# Efficient prediction of potential energy surface and physical properties with Kolmogorov-Arnold Networks

Rui Wang[1,2], Hongyu Yu[1,2], Yang Zhong[1,2], Hongjun Xiang[1,2]

[1] Key Laboratory of Computational Physical Sciences (Ministry of Education), Institute of Computational Physical Sciences, State Key Laboratory of Surface Physics, and Department of Physics, Fudan University, Shanghai, 200433, China
[2] Shanghai Qi Zhi Institute, Shanghai, 200030, China

**Correspondence to:** Prof. Hongjun Xiang, Key Laboratory of Computational Physical Sciences (Ministry of Education), Institute of Computational Physical Sciences, State Key Laboratory of Surface Physics, and Department of Physics, Fudan University, Shanghai, 200433, China. E-mail: hxiang@fudan.edu.cn; ORCID: https://orcid.org/0000-0002-9396-3214

**Abstract**
The application of machine learning methodologies for predicting properties within materials science has garnered significant attention. Among recent advancements, Kolmogorov-Arnold Networks (KANs) have emerged as a promising alternative to traditional Multi-Layer Perceptrons (MLPs). This study evaluates the impact of substituting MLPs with KANs within three established machine learning frameworks: Allegro, Neural Equivariant Interatomic Potentials (NequIP), and the Edge-Based Tensor Prediction Graph Neural Network (ETGNN). Our results demonstrate that the integration of KANs generally yields enhanced prediction accuracies. Specifically, replacing MLPs with KANs in the output blocks leads to notable improvements in accuracy and, in certain scenarios, also results in reduced training times. Furthermore, employing KANs exclusively in the output block facilitates faster inference and improved computational efficiency relative to utilizing KANs throughout the entire model. The selection of an optimal basis function for KANs is found to be contingent upon the particular problem at hand. Our results demonstrate the strong potential of KANs in enhancing machine learning potentials and material property predictions.

**Keywords:** machine learning, property prediction, Kolmogorov-Arnold Networks

## 1. INTRODUCTION

The application of machine learning (ML) methods has become increasingly significant in materials science[1–4]. By leveraging large datasets, ML methods can achieve prediction speeds far exceeding those of traditional methods. For instance, ML techniques offer remarkable accuracy and broad applicability in predicting tensor properties[5], Hamiltonians[6,7], electron-phonon coupling strengths[8], and other properties[9–11] of solids and molecules. Additionally, ML techniques facilitate high-throughput searches for novel materials such as superconductors[12], high-piezoelectric materials[13], Porous Materials[14], and silicon materials[15]. Furthermore, ML potentials for magnetic systems[16,17], metal-organic frameworks[18], and many-body systems[19] demonstrate high accuracy, enabling molecular dynamics simulations over extended time scales[16,18,20,21]. The integration of ML in predicting material properties significantly accelerates the discovery and design of new materials and also enhances our understanding of existing ones.

Multi-layer perceptrons (MLPs)[22,23] are the foundational blocks of most modern machine learning models. Recently, Liu et al. proposed Kolmogorov-Arnold Networks (KANs)[24] as an alternative to

MLPs. KANs are inspired by the Kolmogorov-Arnold representation theorem[25,26]. In KANs, the linear weight parameters are replaced by learnable univariate functions parameterized as splines. Consequently, KANs outperform MLPs in both prediction accuracy and interpretability.

Since the introduction of KANs, numerous variations have been developed by replacing B-splines with different basis functions. Li et al. proposed FastKAN[27], which utilizes radial basis functions (RBFs) with Gaussian kernels, offering a significantly faster implementation of KAN without sacrificing accuracy. Other variations include Wavelet Kolmogorov-Arnold Networks[28] incorporating wavelet functions, Fourier Kolmogorov-Arnold Network for Graph Collaborative Filtering[29], Fractional Kolmogorov-Arnold Network[30] incorporating fractional-orthogonal Jacobi functions, and Kolmogorov-Arnold Networks incorporating sinusoidal basis functions[31]. Additionally, KANs have been applied to a wide range of ML approaches, including Temporal Kolmogorov-Arnold Networks[32] for multi-step time series forecasting, Graph Kolmogorov-Arnold Networks[33] for graph-structured data, and Signature-Weighted Kolmogorov-Arnold Networks[34] using learnable path signatures, among others[35, 36].

Many machine learning models for property prediction rely heavily on MLPs, which makes such models ideal candidates for integrating KANs to enhance prediction accuracy. By substituting MLPs with KANs, it is possible to improve the accuracy of property predictions without modifying the original network architecture. Despite the potential benefits, there has been limited systematic testing in this area. In this study, we investigated the impact of replacing MLPs with KANs on various ML models in property prediction. Specifically, we substituted MLPs in different parts of the machine learning potential Allegro[37] with KANs employing various basis functions. Our results show that replacing the MLPs in the output block of the Allegro model not only enhances prediction accuracy but can also reduce training time in certain cases. Additionally, it improves inference speed and computation resource efficiency relative to using KANs without MLPs. We extended this approach to other models, including Neural Equivariant Interatomic Potentials (NequIP)[38] and the edge-based tensor prediction graph neural network (ETGNN)[5]. Consistently, replacing the MLPs in the output blocks of these models improved prediction accuracy and decreased training time. Overall, using KANs with different basis functions generally enhances prediction accuracy, and the optimal basis function depends on the specific problem. Our findings highlight the significant promise of KANs in enhancing machine learning models for material property prediction and machine learning potentials.

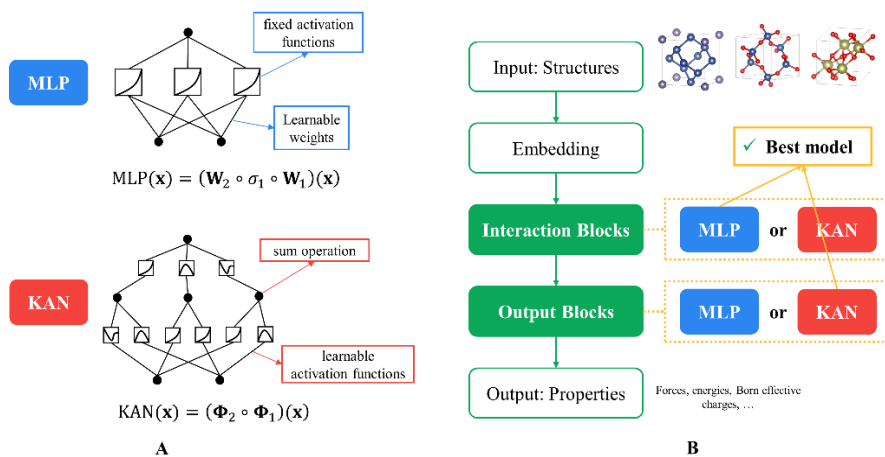

**Figure 1.** Advancing materials property prediction with KAN. A: Comparison of MLP and KAN[24]. MLPs unilize learnable weights on the edges and fixed activation functions on nodes. In contrast, KANs employ learnable activation functions parameterized as various basis functions on edges with sum operations on nodes; B: Replacing MLPs in property prediction models with KANs. The left side illustrates the general framework of property prediction models. In this study, MLPs in different parts of the property prediction models are replaced with KANs employing various basis functions. Our results

demonstrate that replacing MLPs with KANs in the output blocks leads to higher prefiction accuracy and reduced training times compared to using MLPs, and higher inference speed and computation resource efficiency compared to using KANs without MLPs.

## 2. METHODS

### 2.1. Machine learning potential Allegro using KAN

First, we utilized Allegro[37] to assess the impact of replacing MLPs in various parts of machine learning potentials with KAN networks employing different basis functions. Allegro[37] is an equivariant deep-learning interatomic potential. By integrating equivariant message-passing neural networks[39] with strict locality, Allegro achieves high prediction accuracy, generalizes well to out-of-distribution data, and scales effectively to large system sizes.

*2.1.1. Replacing all MLPs with KANs with different basis functions*

First, we tried replacing all MLPs in the Allegro model with KANs using different basis functions. We substituted MLPs in three parts of the Allegro model with KANs: the two-body latent embedding part, the latent MLP part, and the output block, as the second model shown in Figure 2. The function of the two-body latent embedding part is to embed the initial scalar features into the latent features of atom pairs. The latent MLP passes information from the tensor products of the current features to the scalar latent space. The output block predicts pairwise energies using the output from the final layer. We did not replace the MLPs in the environment embedding part, as it typically consists of a simple one-layer linear projection, making it trivial to substitute with KANs.

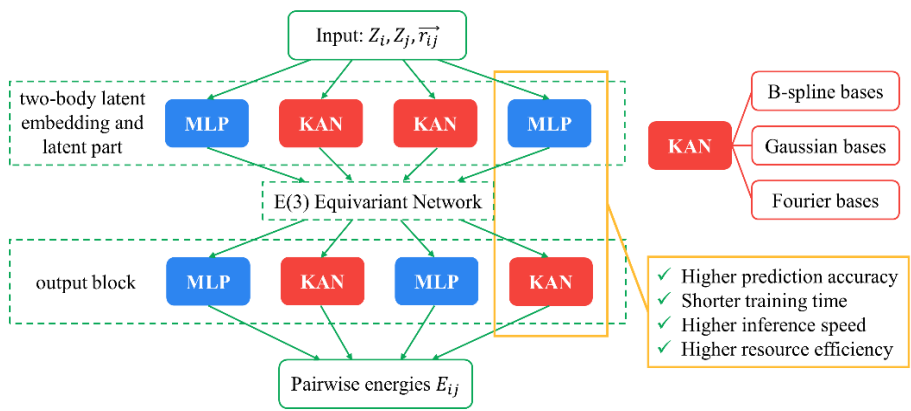

**Figure 2.** Replacing MLPs in different parts of the machine learning potential Allegro[37] with KANs employing various basis functions. $Z_i$ stands for the chemical species of atom $i$. $\vec{r_{ij}}$ stands for the relative displacement vector from atom $i$ to atom $j$. Substituting MLPs with KANs generally enhances prediction accuracy. Specifically, replacing MLPs in the output block of the Allegro model results in higher prediction accuracy and shorter training time than using MLP, and higher inference speed and higher computation resource efficiency than using KANs throughout the entire model.

For the Allegro model utilizing KANs, we tested KANs with the original B-spline basis functions, Gaussian functions, and Fourier functions. For KANs with B-spline basis functions, we employed the efficient-kan package[40], a re-implementation of the original KAN with enhanced efficiency. For KAN implementations with Gaussian and Fourier functions, we used the fastkan package[27]. The details of different models are included in Supplementary Materials.

We evaluated the accuracy and efficiency of various models using the Ag dataset[37]. This dataset was derived from ab-initio molecular dynamics simulations of a bulk face-centered-cubic structure with a vacancy, consisting of 71 atoms. The simulations were performed using the Vienna Ab-Initio Simulation Package (VASP)[41] with the PBE exchange-correlation functional[42]. The dataset includes 1,000 distinct structures, with 950 used for training and 50 used for validation.

*2.1.2. Replacing some of MLP in Allegro with KAN*

We subsequentially replaced MLPs in various parts of the Allegro model with KANs to identify the optimal configuration. We selected KANs with Gaussian basis functions based on the results in section 3.1.2, which provides higher prediction accuracy while maintaining relatively short training times. Specifically, we evaluated two configurations: incorporating KANs in the two-body latent embedding and latent MLP parts and incorporating KANs solely in the output block, as shown in Figure 2. The details of different models are included in Supplementary Materials.

We evaluated the performance of various models on the Ag dataset and $HfO_2$ dataset[43]. The Ag dataset is identical to the one described in the previous section. The $HfO_2$ dataset[43] was generated using density functional theory calculations performed with the VASP package[41]. The structures were initially generated by perturbing ground-state $HfO_2$ structures, followed by sampling through NPT simulations at various temperatures. We selected 10000 structures from the dataset, with 9000 used for training and 1000 used for validation.

We also evaluated the inference speeds and GPU memory usage of various models by performing molecular dynamics simulations using the Large-scale Atomic/Molecular Massively Parallel Simulator (LAMMPS)[44]. The simulations employed the Allegro pair style implemented in the Allegro interface[37]. The initial structure was obtained from the Ag dataset[37]. The simulations were conducted under an NVT ensemble at a temperature of 300K with a time step of 1 ps. For each model, we ran 5,000 time steps to measure the inference speed.

## 2.2. Machine learning potential NequIP using KAN

We also investigated replacing MLPs with KANs in the NequIP model[38], a deep-learning interatomic potential. Neural Equivariant Interatomic Potentials (NequIP) utilize E(3)-equivariant convolutions to capture interactions between geometric tensors, resulting in exceptional prediction accuracy and remarkable data efficiency.

The NequIP architecture is based on an atomic embedding that generates initial features from atomic numbers. This embedding is followed by interaction blocks that integrate interactions between neighboring atoms through self-interactions, convolutions, and concatenations. The final output block converts the output features of the last convolution into atomic potential energy. As with the optimal model in section 3.1.2, we only replaced the MLPs in the output blocks with KANs, as shown in Figure 3. We tested KANs with Gaussian and B-spline bases, utilizing the efficient-kan package[40] for the B-spline bases and the fastkan package[27] for the Gaussian bases. The details of different models are included in Supplementary Materials. We tested NequIP with MLPs and KANs on the Ag dataset identical to the one used in previous sections.

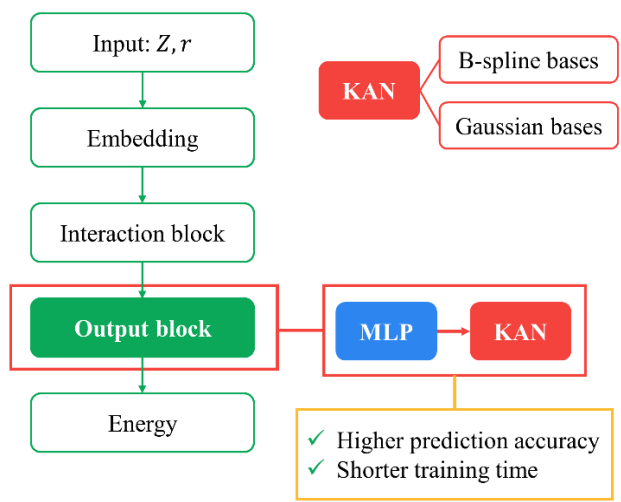

**Figure 3.** Replacing MLPs in the output block of the NeqIP[38] model with KANs employing B-spline and Gaussian basis functions. $e$ and $\alpha$ stand for the lengths of the edges and the angles between the edges in the cluster. Substituting the MLP with the B-spline bases KAN improves prediction accuracy and significantly shortens the training time.

### 2.3. Tensor prediction networks (ETGNN) using KAN

In this study, we utilized the edge-based tensor prediction graph neural network (ETGNN)[5] to predict the tensorial properties of crystals. In ETGNN, tensorial properties are represented by averaging the contributions of atomic tensors within the crystal. The tensor contribution of each atom is decomposed into a linear combination of local spatial components, which are projected onto the edge directions of clusters with varying sizes. This approach enables ETGNN to predict the tensorial properties of crystals with efficiency and accuracy while maintaining equivariance.

In the ETGNN architecture, the initial features are generated in the embedding block and subsequently updated through a series of update blocks. The output of the final update block is then aggregated into node features by the node output block to produce scalar outputs. As represented in Figure 4, consistent with our modifications to the machine learning potentials, we only replaced the projection part of the MLPs from the edge update block and the node output block, which, similar to the output block in machine learning potential models, convert the output features into scalars. We replaced the MLPs with KANs using Gaussian and B-spline bases. The details of different models are included in Supplementary Materials.

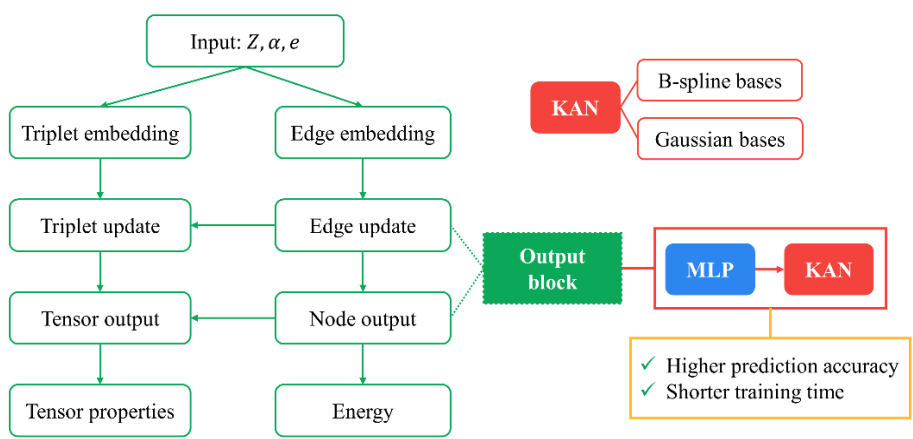

**Figure 4.** Replacing MLPs in the output block of the ETGNN model[5] with KANs employing B-spline and Gaussian basis functions. $e$ and $\alpha$ stand for the lengths of the edges and the angles between the edges in the cluster. Replacing the MLP in the output block with a KAN using Gaussian basis functions significantly improves prediction accuracy while also reducing training time.

We compared the accuracy of ETGNN using MLPs and KANs with different basis functions on a $SiO_2$ dataset[5]. The dataset consists of 3,992 randomly perturbed $SiO_2$ structures calculated using density functional perturbation theory (DFPT). The dataset was split into training, validation, and test sets in a 6:2:2 ratio. We calculated the Born effective charges using ETGNN with MLPs and KANs with Gaussian and B-spline basis functions.

## 3. RESULTS AND DISCUSSION

### 3.1. Allegro using KAN

#### 3.1.1. Replacing all MLP in Allegro with KAN with different basis functions

We evaluated the performance of replacing MLPs in the Allegro model with KANs using various basis functions on the Ag dataset. The mean absolute error (MAE) and training times of the predicted potentials are presented in Table 1 and Figure 5. Notably, all three Allegro models using KANs demonstrated lower force and energy MAE than the original Allegro model with MLPs. Specifically, the force MAE for the KAN-based model with Gaussian bases is 0.014 eV/Å, which is 12.5% lower than that of the MLP-based Allegro model. The model utilizing KANs with B-spline bases achieved the lowest validation energy MAE of 0.029 eV/atom, which is 17.1% lower than the MLP-based model. However, this model required nearly five times the training time. Conversely, the Allegro model with KANs using Gaussian bases also exhibited a lower validation energy MAE than the MLP-based model, 0.032 eV/atom, while maintaining a comparable training time. The model with Fourier bases resulted in a validation energy MAE similar to the MLP-based Allegro model but required a longer training time.

All three Allegro models using KANs demonstrated superior prediction accuracy compared to Allegro using MLPs. This improved performance may be attributed to the fact that basis functions like splines offer better fitting capabilities than MLPs[24,36], providing significant advantages in solving complex problems such as predicting material properties. The Allegro model using B-spline basis functions exhibited significantly longer training time than those using other bases. This is likely due to the substantial computational time required for calculating B-spline parameters[36]. Employing more efficient basis functions, such as Gaussian and Fourier functions, can enhance model efficiency.[24,27].

**Table 1. Results of Allegro with MLPs and KANs with B-spline, Gaussian, and Fourier basis functions. The best results are written in bold.**

| Model | training F MAE (eV/Å) | training E MAE (eV/atom) | validation F MAE (eV/Å) | validation E MAE (eV/atom) | Training Time |
|---|---|---|---|---|---|
| Allegro using MLPs | 0.016 | 0.028 | 0.016 | 0.035 | **4h 51m** |
| Allegro using KAN with B-spline bases | **0.014** | **0.021** | **0.014** | **0.029** | 22h 45m |
| Allegro using KAN with Gaussian bases | **0.014** | 0.026 | **0.014** | 0.032 | 4h 56m |

| Allegro using KAN with Fourier bases | **0.014** | 0.025 | **0.014** | 0.037 | 6h 54m |

MAE: mean absolute error; F: force; E: energy.

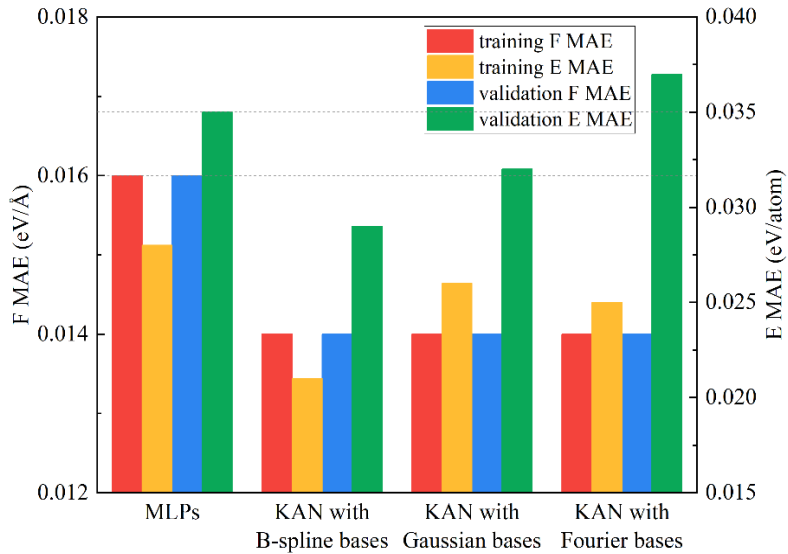

**Figure 5.** The mean absolute error (MAE) of replacing MLPs in the Allegro model with KANs using various basis functions. All three Allegro models using KANs demonstrated lower force and energy MAE than the original Allegro model with MLPs.

### 3.1.2. Replacing some of the MLPs in Allegro with KAN

We subsequentially replaced MLPs in various components of the Allegro model with KANs to identify the optimal configuration. We chose KANs with Gaussian basis function, as they provide higher prediction accuracy while maintaining relatively short training times. Specifically, we evaluated two configurations: incorporating KANs in the two-body latent embedding and latent MLP parts and incorporating KANs solely in the output block.

We initially evaluated the performance of various models using the Ag dataset, with results presented in Table 2 and Figure 6. Remarkably, the Allegro model incorporating KANs in the output block achieved the highest prediction accuracy, a validation energy MAE of 0.022 eV/atom, which is 37.1% lower than that of Allegro using MLPs. The Allegro model utilizing KANs in the two-body latent embedding and latent MLP parts demonstrated slightly improved prediction accuracy and reduced training time compared to the model using MLPs.

**Table 2. Results of replacing MLP from different parts of Allegro with KAN using Gaussian bases on the Ag dataset. The best results are written in bold.**

| Model | training F MAE (eV/Å) | training E MAE (eV/atom) | validation F MAE (eV/Å) | validation E MAE (eV/atom) | Training Time |
|---|---|---|---|---|---|
| Allegro using MLPs | 0.016 | 0.028 | 0.016 | 0.035 | **4h 51m** |
| Allegro using KAN in the | 0.015 | **0.025** | 0.015 | 0.028 | 5h 20m |
| two-body latent embedding and latent MLP part | | | | | |
| Allegro using KAN in the output block | **0.014** | **0.025** | **0.014** | **0.022** | 9h 40m |
| Allegro using KAN without MLP | **0.014** | 0.026 | **0.014** | 0.032 | 4h 56m |

MAE: mean absolute error; F: force; E: energy.

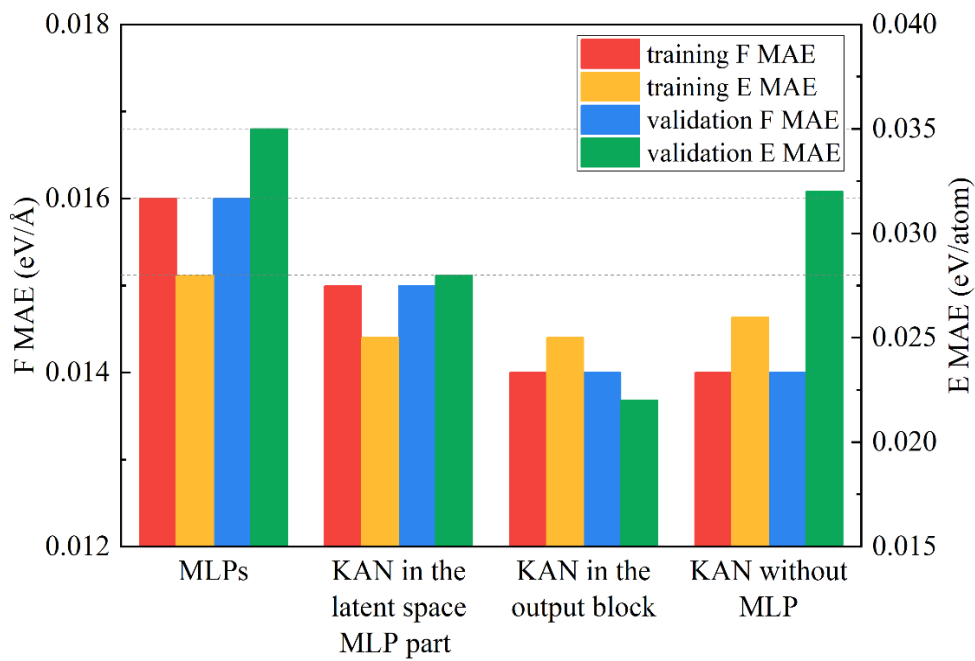

**Figure 6.** The mean absolute error (MAE) of replacing MLPs in various components of the Allegro model with KANs on the Ag dataset. All three Allegro models using KANs demonstrated lower force and energy MAE than the original Allegro model with MLPs. Remarkably, the Allegro model incorporating KANs in the output block achieved the highest prediction accuracy.

The inference speeds and GPU memory usage of various models are shown in Table 3. In general, the Allegro models using KANs with different basis functions exhibited slightly higher GPU memory usage compared to those using MLPs. This suggests that Allegro models employing MLPs are more efficient in terms of model design and data handling, leading to better computation resource efficiency. Replacing only some of the MLPs in the Allegro model with KANs led to a reduction in GPU memory usage. Specifically, the Allegro model with KANs in the output block required 1945 MB GPU memory, just 4 MB more than the 1941 MB used by the Allegro model with MLPs. The inference speed of Allegro using KANs was only slightly slower than that of the model using MLPs. Specifically, the inference speed of the Allegro model with KANs in the output block is 8.92 ms per time step, only 0.70 ms per time step slower than the Allegro model using MLPs. Using KAN solely in the output block improves prediction accuracy compared to using MLP, and also improves inference speed and computation resource efficiency compared to using KANs throughout the entire Allegro model.

**Table 3. The inference speed and GPU memory usage of replacing MLP from different part of Allegro with KAN using Gaussian bases.**

| Model | inference speed (ms per time step) | GPU memory usages (MB) |
|---|---|---|
| Allegro using MLPs | 8.22 | 1941 |
| Allegro using KAN in the two-body latent embedding and latent MLP part | 9.24 | 1963 |
| Allegro using KAN in the output block | 8.92 | 1945 |
| Allegro using KAN without MLP | 9.44 | 1963 |

The improvements observed in the results on the Ag dataset were modest, likely because the dataset is too simple to benefit significantly from KANs[24]. Therefore, we proceeded to evaluate these models on the more complex $HfO_2$ structures.

The results, as presented in Table 4 and Figure 7, demonstrate that replacing the MLP in the output block of Allegro significantly improves prediction accuracy for both energies and forces. The validation force MAE is reduced to 0.054 eV/Å, a decrease of 27.0% compared to Allegro with MLPs. Similarly, the validation energy MAE is reduced to 0.104 eV/atom, which is 36.6% lower than with MLPs. Additionally, the training time is notably shortened. Furthermore, the GPU memory allocated during the training process of the Allegro model using KAN in the output block is 45.63%, only 0.03% higher than using MLP. However, replacing MLPs in other parts of the Allegro model has minimal impact on either prediction accuracy or training time. These findings are consistent with the results obtained from the Ag dataset.

**Table 4. Results of replacing MLP from different parts of Allegro with KAN using Gaussian bases on the $HfO_2$ dataset. The best results are written in bold.**

| Model | training F MAE (eV/Å) | training E MAE (eV/atom) | validation F MAE (eV/Å) | validation E MAE (eV/atom) | Training Time |
|---|---|---|---|---|---|
| Allegro using MLPs | 0.076 | 0.265 | 0.074 | 0.164 | 7d 3m |
| Allegro using KAN in the two-body latent embedding and latent MLP part | 0.064 | 0.473 | 0.063 | 0.172 | 7d 2m |
| Allegro using KAN in the output block | **0.053** | **0.146** | **0.054** | **0.104** | **4d 11h 40m** |
| Allegro using KAN without MLP | 0.058 | 1.444 | 0.056 | 0.200 | 7d 10m |

MAE: mean absolute error; F: force; E: energy.

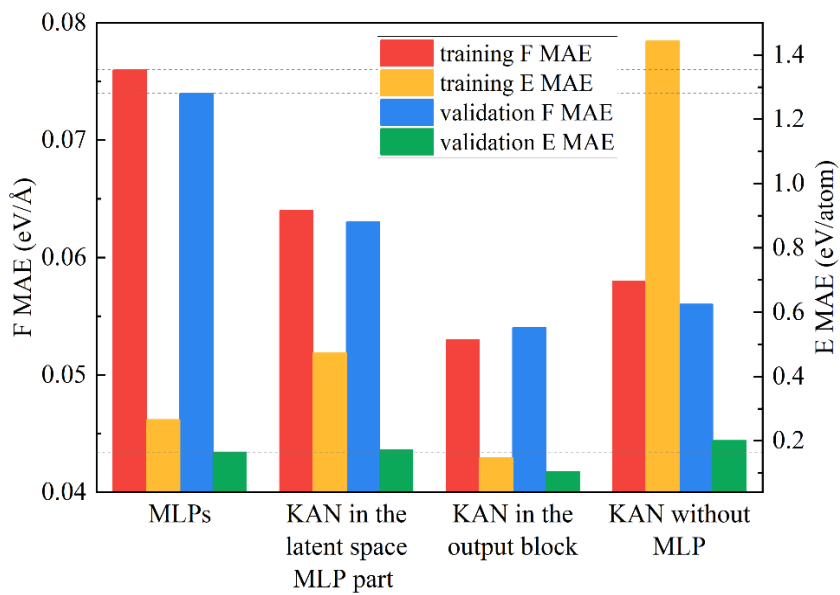

**Figure 7**. The mean absolute error (MAE) of replacing MLPs in various components of the Allegro model with KANs on the $HfO_2$ dataset. Replacing the MLP in the output block of Allegro significantly improves prediction accuracy for both energies and forces.

Replacing the MLP in the output block of the Allegro model with KAN significantly improves prediction accuracy. In some cases, this substitution also reduces training time. This improvement occurs because KANs are more effective at fitting functions[24,36]. However, basis functions like splines are less capable of exploiting compositional structures and therefore inferior to MLPs in feature learning[24]. Consequently, the output block, which predicts energies from the final layer's output, is well-suited for KANs to enhance prediction accuracy. Therefore, using KANs in other parts of the Allegro model, such as the embedding layer, results in smaller improvements in prediction accuracy compared to the output block.

### 3.2. Nequip using KAN

Based on the findings from the previous section, we replaced the MLPs in the output block of the NequIP model with KANs. We tested NequIP with MLPs and KANs with Gaussian and B-spline bases on the Ag dataset. The results are shown in Table 5. All three models exhibited similar accuracy, likely due to the simplicity of the Ag dataset[24,36]. Additionally, replacing the MLP with the Gaussian bases KAN did not reduce training time. However, substituting the MLP with the B-spline bases KAN significantly shortened the training time.

**Table 5. Results of replacing MLP from the output block of NequIP with KAN using Gaussian bases and B-spline bases on the Ag dataset. The best results are written in bold.**

| Model | training F MAE (eV/Å) | training E MAE (eV/atom) | validation F MAE (eV/Å) | validation E MAE (eV/atom) | Training Time |
|---|---|---|---|---|---|
| NequIP using MLPs | 0.011 | 0.015 | 0.013 | 0.015 | 2d 8h 55m |
| NequIP using KAN with Gaussian bases in the output block | 0.011 | 0.015 | 0.013 | 0.015 | 2d 11h 46m |

| Model | | | | |
|---|---|---|---|---|
| NequIP using KAN with B-spline bases in the output block | 0.011 | 0.016 | 0.013 | **0.013** | **1d 13h 2m** |

NequIP: Neural Equivariant Interatomic Potentials; MAE: mean absolute error; F: force; E: energy.

### 3.3. Tensor prediction networks (ETGNN) using KAN

We calculated the Born effective charges for the SiO2 dataset using ETGNN models with MLPs and KANs with Gaussian and B-spline basis functions. The results are shown in Table 6. Replacing the MLP in the output block with a KAN using a Gaussian bases significantly improves prediction accuracy while also reducing training time. The result is consistent with what we achieved on the Allegro model. The training time for ETGNN with KANs using B-spline bases is shorter than with MLPs, and longer compared to using KANs with Gaussian bases.

**Table 6. Results of replacing MLP from the output block of ETGNN with KAN using Gaussian bases and B-spline bases on the SiO2 dataset. The best results are written in bold.**

| Model | training MAE (e) | validation MAE (e) | test MAE (e) | Training Time |
|---|---|---|---|---|
| ETGNN using MLPs | 0.00452 | 0.00517 | 0.00502 | 2h 55m |
| ETGNN using KAN with Gaussian bases in the output block | **0.00439** | **0.00473** | **0.00450** | **1h 36m** |
| ETGNN using KAN with B-spline bases in the output block | 0.00547 | 0.00564 | 0.00542 | 1h 51m |

ETGNN: edge-based tensor prediction graph neural network; MAE: mean absolute error; F: force; E: energy.

## 4. CONCLUSIONS

In this study, we assessed the impact of replacing MLPs with KANs in various machine learning models for material property prediction. We found that substituting MLPs with KANs generally enhances prediction accuracy. Specifically, replacing MLPs in the output block of the machine learning model significantly improves accuracy and, in some instances, reduces training time. Moreover, using KANs exclusively in the output block increases inference speed and computation resource efficiency compared to using KANs without MLPs in the property prediction model. The choice of the optimal basis function for KANs depends on the specific problem. Our results demonstrate the strong potential of KANs in machine learning models for machine learning potentials and material property prediction. Relying on KAN's powerful expressive capabilities, Universal MLPs such as M3GNet[45], CHGNet[46], and MACE[47] could be able to achieve better accuracy and generalization.

## DECLARATIONS

### Authors' contributions

Rui Wang and Hongyu Yu contributed equally to this work.
Made substantial contributions to the conception and design of the study and performed data analysis and interpretation: Rui Wang, Hongyu Yu;
Provided administrative, technical, and material support: Yang Zhong, Hongjun Xiang.

**Financial support and sponsorship**

**Conflicts of interest**

All authors declared that there are no conflicts of interest.